# Non-equilibrium Strain Relaxation Noise in the Relaxor Ferroelectric

# $(PbMg_{1/3}Nb_{2/3}O_3)_{1-x}(PbTiO_3)_x$

Xinyang Zhang[1], Thomas J. Kennedy[1], Eugene V. Colla[1], M. B. Weissman[1], and

D. D. Viehland[2]

[1] Department of Physics

University of Illinois at Urbana-Champaign

Urbana, Illinois, 61801, USA

[2] Department of Materials Science and Engineering,

Virginia Tech,

Blacksburg, Virginia, 24061, USA

**Abstract:**

Large low-frequency polarization noise is found in some perovskite relaxor ferreoelectrics when they are partially polarized, regardless of whether the polarization is accompanied by an applied electric field. The noise appears both in the ferroelectric and relaxor states, including in the nominally ergodic paraelectric state at temperatures above the susceptibility peak. Since it is present whenever the samples have non-zero average piezoelectric coefficients, but not otherwise evident, it appears to be a response to mechanical strain changes. Dependence of the noise on sample thermal history indicates that non-equilibrium strain relaxation is the source, even in the temperature range for which the sample is nominally ergodic. Non-equilibrium noise in the absence of net piezoelectricity is found at somewhat higher frequencies. Related materials lacking a metastable non-equilibrium cubic bulk phase and a symmetry-broken surface layer show very little of the anomalous low-frequency noise. The implications for a non-equilibrium origin of the skin effect are discussed.

**Introduction**

Several perovskite relaxor ferroelectrics[1,2] in the PMN-PT family, $(PbMg_{1/3}Nb_{2/3}O_3)_{1-x}(PbTiO_3)_x$ have been observed to show low-frequency polarization noise of unknown origins so prominent that it shows up in pyroelectric current experiments not intended to look for noise.[3,4,5] Since this noise is far above the fluctuation-dissipation noise level required by thermodynamics, it can limit the materials' use as sensitive electromechanical transducers. Unlike the ordinary Barkhausen noise often observed in ferroelectrics[6,7], it usually persists for days after changes in electric field and shows no apparent connection to the net rate of polarization change but instead is dependent on the polarization itself.[3,4,5] Unlike the thermal polarization noise found in some ferroelectrics[8], it is not limited to periods in which a phase transition is taking place. The effect is far more dramatic than the relatively subtle violations of the fluctuation-dissipation relation found in non-equilibrium spinglasses.[9] Its origin presents an interesting puzzle.

It has previously been noted that the noise is present whenever a low-x sample (which we shall call here PMNPTx, where x is the percent rather than the fraction), is polarized and absent when the sample is not polarized.[4,5] Below about x=0.20, bulk samples of these materials exhibit inversion symmetry and thus lack a net spontaneous piezoelectric coefficient[10], acquiring one only when their symmetry has been broken via an applied field or, at low temperature, via being polarized by prior application of a field. Thus they are microphonic, i.e. generate voltages in response to long-range coherent strain only when polarized. Since the noise appears only when the sample is microphonic, extraneous mechanical vibrations are the obvious first suspect for the source, but these have been consistently ruled out by simple tests to reduce or increase such vibrations, e.g. turning off noisy pumps or stomping around near the cryostat.[5] In a previous paper focused on Barkhausen noise, we presented preliminary speculation that this extra low-frequency noise was internally generated, i.e. that the material is mechanically creaking due to very slow thermal equilibration.[5]

Here we present evidence, based largely on sensitivity to thermal and field histories and on the persistence of the noise in the non-ferroelectric state, confirming that speculation. We also find that the anomalous low-frequency noise is almost absent in higher-x materials lacking the bulk cubic so-called "X phase" and the broken-symmetry "skin effect" surface layer that often accompanies the X phase.[11-15] We shall discuss possible implications of the low frequency noise for the origin of the poorly understood skin effect and the metastability of the X phase.

**Materials and Methods**

The samples of PMNPTx are from the same batch previously described in our work on the kinetics and thermodynamics of forming the ferroelectric (FE) phase.[16] They were grown by a modified Bridgman technique and supplied by TRS Technologies (State College, PA). Fig. 1 shows an empirical phase diagram of a PMNPT12 sample from this batch, illustrating the ergodic paraelectric (PE), non-ergodic relaxor paraelectric (RXF), and the FE regimes, as well as the hysteresis between RXF and FE. The melting temperature of the FE phase is 302K. The PMNPT20 sample shows similar behavior with somewhat higher characteristic temperatures, with a FE melting at 341K. The noise studies here included the FE regime, the metastable RXF regime, the stable RXF regime, and the PE regime. The PMNPT32 sample goes directly from a PE phase to an FE domain phase at roughly 430 K, but its FE phase depolarizes at ~411 K. The noise studies here focus on its transition range.

As background, throughout the temperature range of our measurements these materials are largely filled with polar nanodomains (PNDs), on a scale of ~10nm, each locally piezoelectric but with quasi-random orientations.[17-19] Only the FE phase has bulk long-range polar order. The bulk of the RXF phase is referred to as "phase X", with overall non-polar cubic symmetry.[11-15] Within about 100µm of the surface, a lower-symmetry skin-effect region is observed.[11-15] A comprehensive discussion of the evidence for phase X and the skin effect may be found in reference [15], with detailed results on phases of powder samples in ref [20].

The samples were configured as parallel-plate capacitors oriented with the applied **E** along a [111] axis, an easy polarization axis for the ferroelectric state. Contacts were made via evaporated Ag layers of roughly 200nm thickness on top of adhesion-enhancing ~10 nm thick evaporated Cr layers. The PMNPT12 sample started at ~0.4mm thick, becoming slightly thinner for later measurements after repolishing. Its contact area is a 1 mm$^2$ disk out of a crystal area ~2 mm by ~3 mm. The PMNPT20 sample is 0.48mm thick with contacts 1.11 mm by 0.75mm out of a total area ~4.22mm by ~0.75mm. The PMNPT32 sample is 0.44mm thick with a total contact area 2.8mm² out of an irregular shape several times larger.

The measurement circuitry, described elsewhere[5], allows the sample voltage to be fixed with ac and dc biases while the polarization current $I_P(t)$ is measured at a rate of ~10 samples/sec. Using a low-pass filter (usually set at 20 Hz) allows simultaneous measurement of the systematic polarization current and low-frequency (LF) noise in $I_P(t)$ along with the complex dielectric response function, ε'-iε", measured at 100 Hz (using a 28.3 mV rms ac drive) on an unfiltered channel.[5] High-frequency (HF) noise was measured via a channel with a high-pass filter set at 1.0 Hz to remove the main large deterministic signal followed by an anti-alias low-pass filter at 2.1 kHz with 5kHz sampling, in runs where the ac susceptibility was not measured.[5] In order to reduce electromagnetic pickup the nitrogen-flow transfer-line cryostat was mounted inside a double-wall mu-metal shield. The shield was supported on a sand pile to reduce vibrational pickup.

**Results**

As before [3-5], we found large low-frequency noise when the low-x samples were polarized. The noise was insensitive to very large changes in acoustic input, including turning the nearby vacuum pump on or off and even tapping softly on the transfer line. Fig. 2 shows $I_P(t)$ measured at 250K in an applied dc voltage, as the PMNPT12 sample converts from the metastable RXF phase to the FE phase over several hundred seconds. The noise in $I_P(t)$ grows steadily as the polarization creeps up in the RXF phase, then grows further as the polarization increases in the fairly abrupt transition to the FE phase. After the transition the noise gradually decreases slightly. When the applied voltage is removed after conversion to the FE state, causing little change in the polarization, the noise magnitude remains approximately constant. As the sample is subsequently heated the noise magnitude changes as a function of T, reaching a minimum at about 280K, rising as the sample starts to depolarize, then falling to very low levels after the sample abruptly depolarizes on the transition back to the relaxor phase. The noise level and its dependence on polarization during this sort of protocol were very similar to those observed in PMN some years ago. [3]

The large noise during the melting transition may be viewed as a type of thermal Barkhausen noise.[8] The persistent large noise as the polarized sample sits under fixed conditions before warming does not resemble any conventional Barkhausen noise.[6, 7]

Figure 3 shows similar data for the PMNPT20 sample. LF noise again is found in the polarized sample, but its level is significantly lower than in the similar PMNPT12 or PMN samples, roughly three orders of magnitude in spectral density. (We have seen a similar reduction in each 20% sample studied.) For PMNPT20, unlike for PMNPT12, the LF noise reaches approximately its maximum amplitude during the polarization creep phase before the FE transition, with about the same amplitude after the transition.

Figure 4 shows data for the PMNPT32 sample, which forms an aligned FE state almost immediately when a field of 550V/cm is applied. The sample shows thermal Barkhausen noise (toward the left side of the figure) as it is cooled at E=0 through the transition temperatures expected for the two FE phase changes [15] at this composition. The resulting state is FE but lacks overall polarization due to since the domains are not systematically aligned. After field application, while sitting at fixed T in the polarized FE state, the noise is substantially lower than in even the PMNPT20 sample, i.e. barely above instrumental background except for an occasional Barkhausen spike. Since the contact area is larger than for the other samples, the current noise, which is additive over area, is comparatively even lower per contact area.

To test whether the low-frequency noise inherently involved the FE phase we looked for it above the equilibrium FE transition line, in the PE state of PMNPT12. The LF noise again appeared, as shown in Fig. 5, when the sample was polarized via an applied voltage. We cannot separately check the voltage and polarization dependences in this regime, in which the polarization relaxes back to very nearly zero after the applied field is removed, unlike in the lower-temperature regime, which exhibits large remnant FE polarization. The presence of the LF noise in this regime, above the peak in $\varepsilon'(T)$, shows that it does not require either any long-range FE order or even the major slow polarization response of the non-ergodic RXF regime. Nevertheless, the asymmetry (more upward spikes) evident in the plots indicates that despite being in the PE regime, a small amount of the sample polarization is occurring after long delays via occasional steps.

Immediately after the voltage is applied in this regime, there is a transient period in which the LF noise gradually builds up. That effect is not simply due to an increase in the microphonic sensitivity, since the polarization itself does not show a significant delayed build-up. On longer time scales, as shown in Fig. 6, the LF noise sometimes starts to decrease again, consistent with it coming from very slow relaxation toward equilibrium.

Since the noise appears to come from some microphonic sensitivity to slow strain relaxations in response to thermal and field history, we checked whether it could be reduced by annealing the sample at high temperature, 773 K, to reduce internal strain. On the initial cool-down after such annealing, the LF noise in the PMNPT12 was indeed substantially reduced, although not eliminated. The insert of Fig. 6 shows the most dramatic effect of this treatment, that on the first post-anneal warming after cooling there is a temperature range around 200 K - 250 K in which the noise magnitude is much lower than it had been on similar warming prior to the annealing. This result is consistent with annealing reducing the non-equilibrium strain, but could also be consistent with other explanations, since the contacts had to be re-applied after the annealing. After further thermal cycling and field application, the LF noise returned to approximately the pre-annealing magnitude, although with a slightly shifted temperature of the minimum on warming. Fig. 6 also shows a slow decrease in noise magnitude during a long period at 150K before warming.

To account for why the low-frequency noise becomes evident only when the samples have net polarization, the low-frequency strain changes must be correlated over regions containing many PND. That would give a piezoelectric voltage that is a coherent sum when the sample is polarized but a much smaller incoherent sum in the net unpolarized condition. One would then expect there to be some non-equilibrium polarization noise even in unpolarized samples, although not easy to pick up against background instrumental noise at low frequencies.

The PMNPT20 sample, which shows relatively little LF noise, gave non-equilibrium noise in the HF channel even while sitting at 370 K with E=0 after cooling from 600 K anneal. This temperature is approximately where the 100 Hz susceptibility peaks. Typical spectra are shown in Fig. 7. The noise magnitude gradually decreased with a typical time-scale of hours. The approximate absolute magnitudes and general time course were reproducible. The susceptibility also showed aging under these conditions, but the magnitude of the susceptibility aging was much too small to account for the size of the noise aging. Other experiments, not shown, found spectral density a little higher than the 9 hour results as the sample passed through the range 365 K to 375 K on rapid warming of the sample from 350 K. Although this extra noise, like ordinary thermal Barkhausen noise[8], is caused by changing temperature, in this material it is found well above the temperature at which long-range FE order melts.

**Discussion**

Polarized PMN and low-x PMNx show low-frequency current noise far above the value expected in equilibrium while held at fixed E and T. The magnitude depends strongly on sample thermal and field history. Although after temperature cycling and field changes the noise persists for periods of days or longer, it does at least sometimes gradually reduce, as expected for a non-equilibrium effect. Even non-polarized PMNPT20 in the PE state can show significant non-equilibrium current noise in a frequency range comparable to the typical dielectric relaxation rate. The dependences of the excess noise on field and temperature and on their histories are consistent with a picture of non-equilibrium strain relaxation. Even when a sample appears to be near equilibrium, i.e. with its polarization very close to the long-term expectation for a given average field and temperature, the internal pattern of the PND's, interacting both by strain and electric fields, appears usually to be far from equilibrium.

Although unpolarized samples do not show the large excess low-frequency noise, they can show a slowly decreasing non-equilibrium noise at higher frequencies. We believe that the explanation for the distinction is that the long-range correlated relaxations are slow and show up very little unless there is systematic piezoelectricity, while faster strain relaxations with only short-range correlations show up about equally regardless of whether the piezo coefficients of different PND have the same sign. The virtual absence of the excess low-frequency noise in unpolarized low-x samples shows that it comes from strain changes that are coherent over distances large compared to the size of polar nanodomains.

That the LF noise is most prominent in the low-x samples, for which the X-phase is most stable[15], suggests a connection to that phase, which is out of thermal equilibrium[16] below the melting line shown on Fig. 1. Since our samples are only several times the thickness of the temperature-dependent strained skin layer[11-15], the existence of major temperature-dependent strains is not a surprise. What static measurements had not shown, however, is that such strains remain out of equilibrium for very long times, with their slow approach to equilibrium creating dramatic polarization fluctuations, even in the PE phase or in the FE phase.

The existence of this noise may provide a clue to one unsolved question about the skin layer, i.e. the origin of the ~100μm characteristic length scale, some $10^4$ times the typical PND length scale. The noise indicates that the strain is far from static, so the skin depth may not be an equilibrium property but rather a scale set by the slow kinetics of long-range strain changes in phase X under typical experimental conditions. The symmetry-broken skin-effect region would be, if this interpretation is correct, closer to equilibrium than is the X phase, with the growth of the skin-effect region inhibited by the slow kinetics of large-scale strain changes in the X phase. It would be interesting to explore whether the low-frequency non-equilibrium noise effects are consistent with some of the simpler pictures of the relaxor state, e.g. reference [21].

The large long-lasting non-equilibrium noise presents obvious difficulties for use of the low-x relaxor materials as sensitive low-noise strain detectors. Its magnitude does depend on field and temperature history and does decrease (sometimes very gradually) under constant E-T conditions, so the difficulties should not be insurmountable.

**Acknowledgments:** X. Zhang was funded by the John A. Gardner Undergraduate Research Award. TJK was supported by the Lorella M. Jones Summer Research Award and the Philip J. and Betty M. Anthony Undergraduate Research Award. We are particularly grateful to a referee who prompted us to explore the relation to phase X and the skin effect.

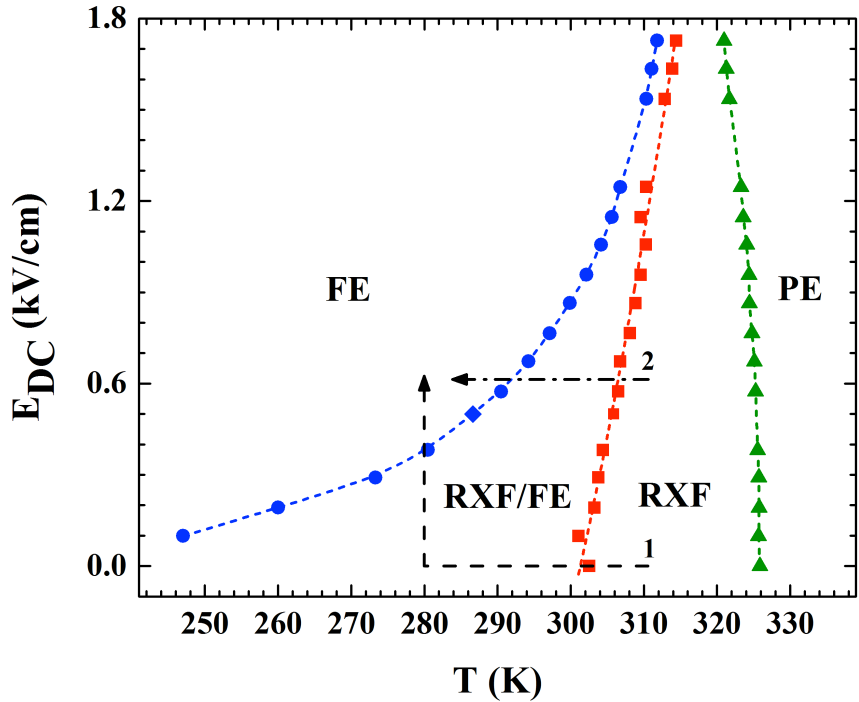

Figure 1. The empirical phase diagram of a PMNPT12 sample from the same batch as those used for the noise studies shows the equilibrium paraelectric phase (PE), separated by a frequency-dependent crossover from the non-equilibrium relaxor state (RXF). The RXF is separated from the FE state by a nearly warming-rate-independent melting line and a cooling-rate-dependent freezing line. The RXF/PE crossover points are the temperatures of the 100 Hz susceptibility maxima on cooling in fixed field. The points between RXF/FE and RXF are the temperatures at which the FE order melts on warming at fixed field, as determined by the sharp spike in pyroelectric current. The freezing points between FE and RXF/FE are the temperatures at which the FE order forms on cooling at a rate of 4K/min at fixed field, as determined by the sharp spike in pyroelectric current. Path "1" shows a ZFC process and path "2" shows an FC process crossing into the FE state.

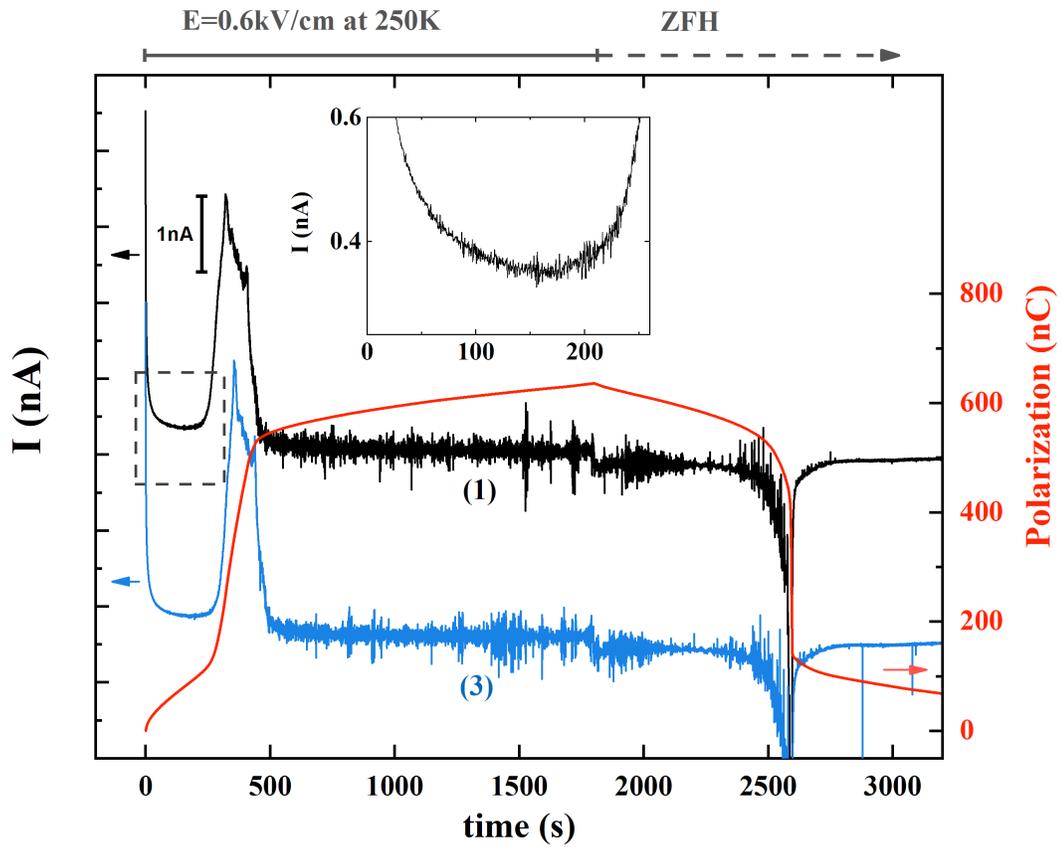

Fig. 2 Shows typical plots of the low-frequency $I_P(t)$ during polarization and depolarization of a PMNPT12 sample. The experimental procedure was: 1) ZFC at 4 K/min to 250 K, aging at 250 K with E= 0.6 kV/cm applied at t=0 for 1800 s, E set to zero at t= 1800s and starting heating at 4K/min to 450 K; 2) Immediately cool at 4K/min down to 200K. Age at 200 K in 0.6 kV/cm for 12 hrs, ZFH to 450 K; 3) Immediate repeat of process (1). The smooth curve shows the net polarization, rising until the field is turned off at the start of the ZFH step. The noisy curves show the low-frequency $I_P(t)$ for processes (1) and (3). Ferroelectric order forms in the peak near t= 400 s and melts abruptly at t = ~2580s, with T near 302K. The time to the ferroelectric transition after field application was slightly shorter in the first run and the temperature with minimum LF noise magnitude on warming occurred at slightly higher T in that run. The inset shows an expanded view of the noise rise as the sample starts to polarize in the first run.

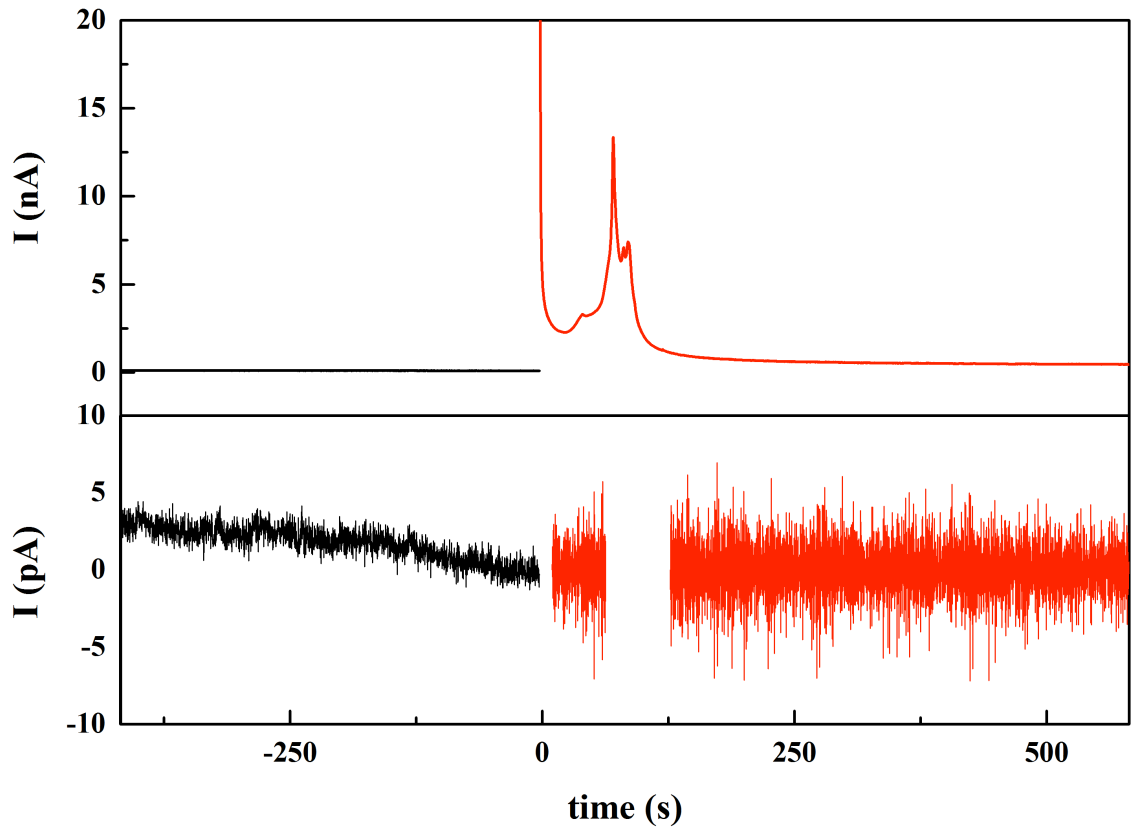

Fig. 3: $I_P(t)$ is shown for the PMNPT20 sample at 295 K, starting with E=0 and with a field of 520 V/cm turned on at t=0. The FE order in this sample melts at 341K. (The origins of the multi-stage melting peak are discussed elsewhere.[5]) The lower graph shows a blow-up of the top curve, with the part after the field was applied de-trended to keep it in-range. The downward trend in the E=0 part at negative times is within the range of LF amplifier drift.

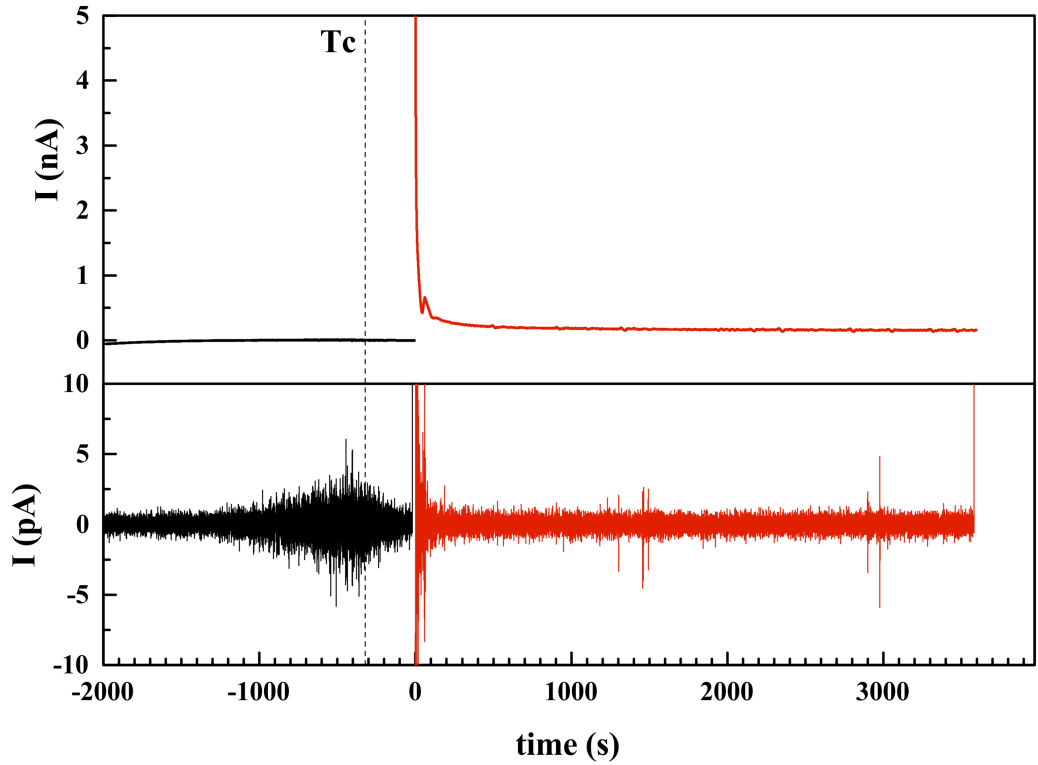

Fig. 4: $I_P(t)$ and a blow-up of detrended $I_P(t)$ for a PMNPT32 sample. At negative times, the sample was cooled in E=0 at 4K/min. At t=0 it reached the target temperature of 395K, at which it was subsequently held, and a field of 550 V/cm was turned on. The FE melting temperature for this sample was 411K, indicated by the dashed line, so the thermal Barkhausen noise mainly occurs in the phase for which the stable phase has FE domains but not systematically aligned.

Zhang 17

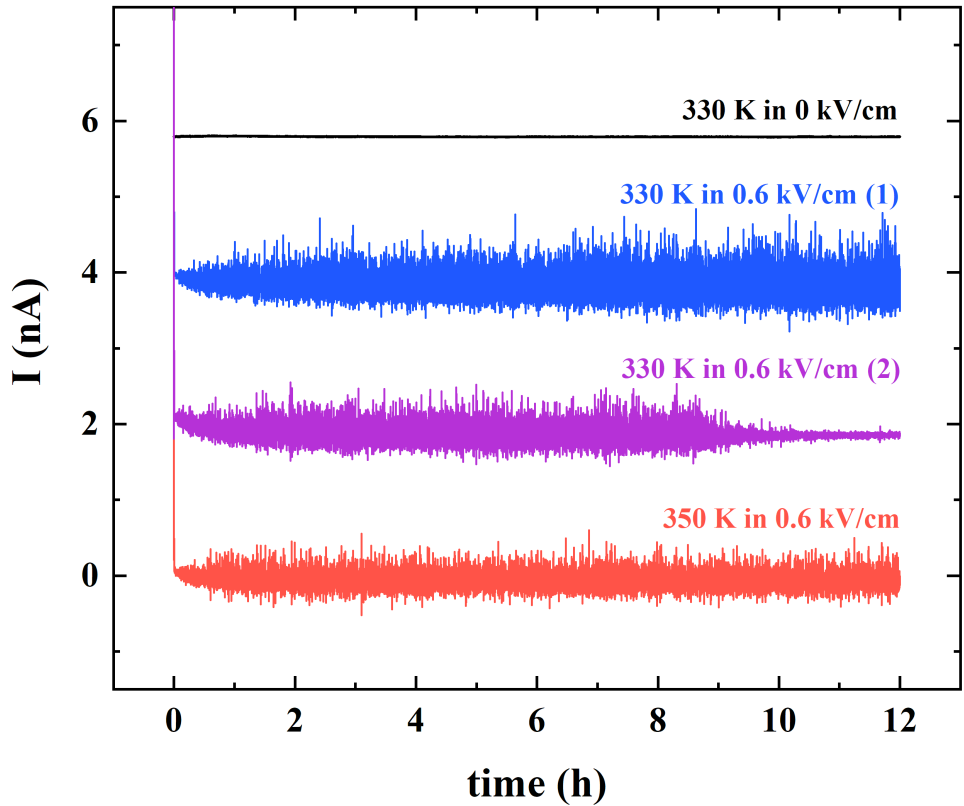

Fig. 5. $I_P(t)$ is shown as a function of time elapsed after applying a field of 0.6 kV/cm to PMNPT12 at 330K or 350K, above the peak in susceptibility vs. T. The top three curves are vertically offset for viewing clarity. Most runs showed LF noise gradually building up over the first hour and remaining about the same magnitude thereafter. However, during one 330K run LF noise fell off dramatically after 10 hrs.

Zhang   18

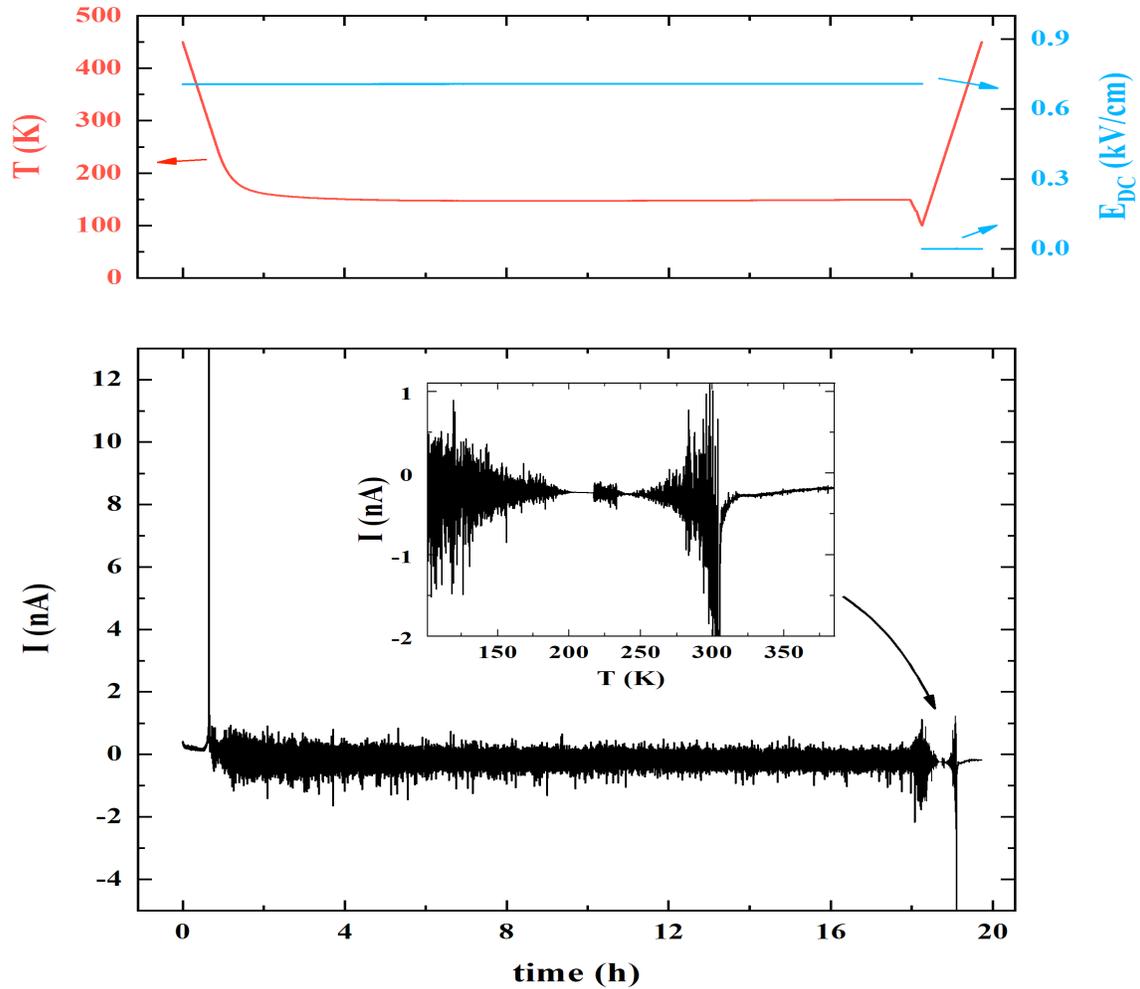

Fig. 6 shows $I_P(t)$ after the PMNPT12 sample was annealed in a furnace in air at 773 K. The temperature ramp rate was roughly 4 K/min for heating and 1 K/min for cooling. Since original electrodes disappeared after the annealing, a new pair of electrodes were deposited right after the annealing. During the initial field-cooling from 450 K to 100 K, the measurement system unexpectedly stuck at ~150 K in 0.7 kV/cm overnight. A ZFH to 450 K followed the unexpected aging at 150 K. As shown in the temperature and field profiles, the field was then set to zero, the sample cooled to 100 K and then warmed in zero field. The LF noise was lower than normal and, as shown in the inset, became negligible in a narrow temperature range during warming.

Zhang	19

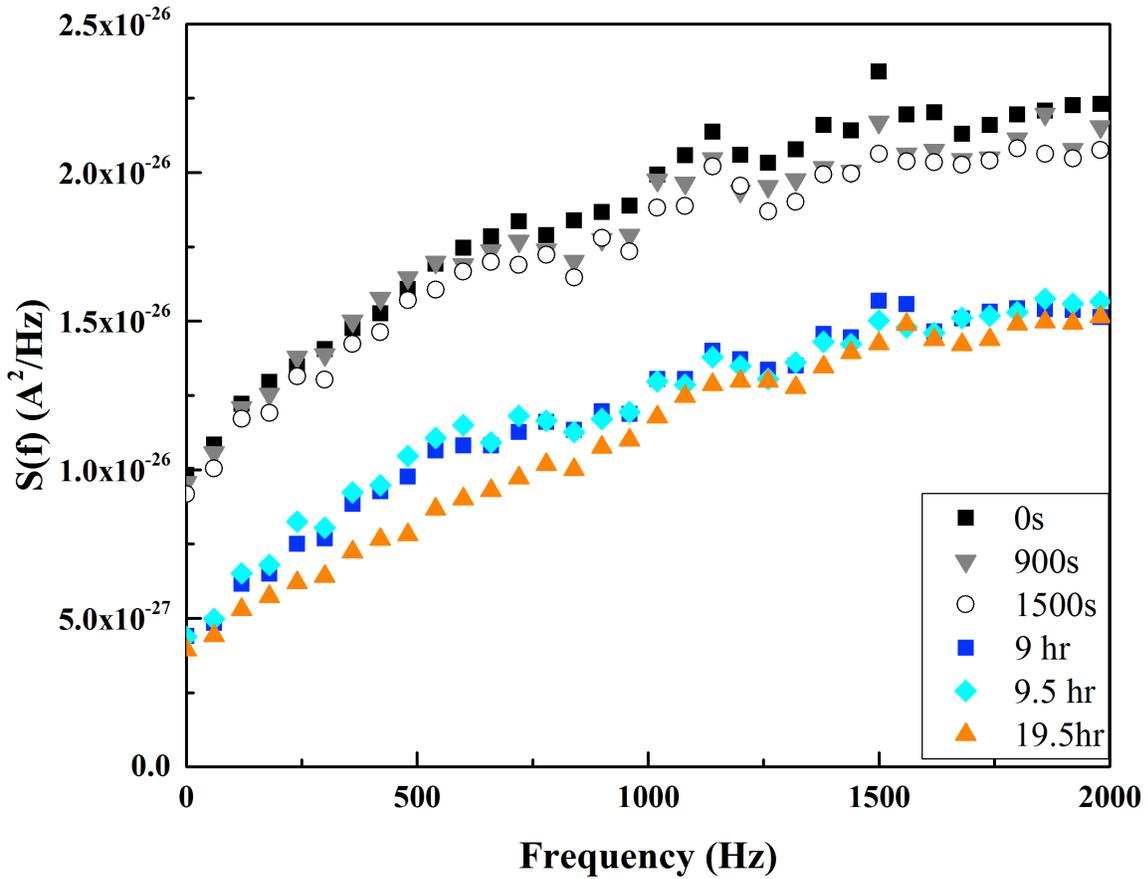

Fig. 7 shows noise spectra taken for the PMNPT20 sample as it sat at 370K with E=0 after cooling at E=0 from 600K. Each spectrum is taken from a 120 s period starting at the delay times listed after the temperature became stable. The spectral density fell during aging, with most of the drop occurring in the first 9 hours. Instrumental background has not been subtracted.